\documentstyle[aps,prb,multicol,epsf]{revtex}
\begin{document} 
\draft
\renewcommand{\narrowtext}{\begin{multicols}{2} \global\columnwidth20.5pc}
\renewcommand{\widetext}{\end{multicols} \global\columnwidth42.5pc}
\multicolsep = 8pt plus 4pt minus 3pt 

\draft

\title{A study of nonlocal conductivity in high-temperature
superconductors}

\author{S. J. Phillipson and M. A. Moore}
\address{Department of Physics and Astronomy, University of
Manchester, Manchester, M13 9PL, UK.}

\author{T. Blum}
\address{Department of Physics, University of Virginia, 
Charlottesville, VA 22901.}

\date{\today} 
\maketitle

\begin{abstract}
We examine nonlocal conductivity in high-temperature superconductors
from a phenomenological point of view.  
One wants to deduce the properties of the conductivity, especially
its inherent length scales, from the transport data. 
Although this is a challenging {\it inverse} problem, complicated 
further by the experimental data not being completely self-consistent, 
we have made some progress. 
We find that if a certain form for the conductivity is postulated 
then one requires positive ``viscosity" coefficients to reproduce 
some of the experimental results. 
We are able to show that the effects of surfaces on the conductivity 
are likely to be important and draw comparisons with the treatment of 
the surface within the hydrodynamic approach put forth by Huse and 
Majumdar. 
We also develop an approximation scheme for the conductivity which 
is more robust than the hydrodynamic one, since it is stable for 
both positive and negative viscosity coefficients, and discuss the 
results obtained using it.
\end{abstract}


\narrowtext

\section{Introduction} 
\label{intro}

\begin{figure}[tbp]
\centerline{\epsfxsize=7.0cm 
\epsfbox{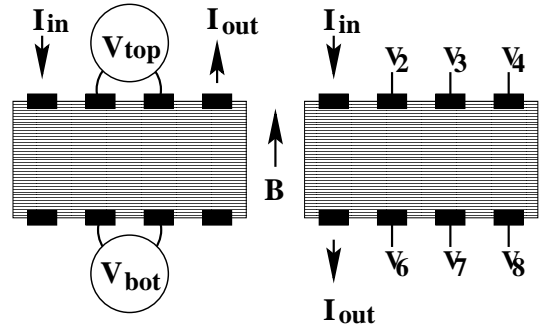}}
\vspace{1cm}
\caption{\label{fluxgeom1}The modified flux-transformer set-up of
Safar {\it et al.}\,\protect{\cite{safar}} 
The diagram on the left shows the arrangement of the terminals for
the {\it top} geometry, and the right shows the {\it side} geometry. 
The samples are single crystals of YBCO with the magnetic field
aligned along the $\hat{c}$-axis of the crystal. 
$V_{left} = V_{2} - V_{6}$ and $V_{right} = V_{3}-V_{7}$.}
\end{figure}

The measurement of a substantial nonlocal conductivity in 
high-temperature superconductors in a magnetic field is thought 
to imply the existence of moving vortex lines having coherence 
lengths of the order of the sample thickness, as opposed to 
pancake vortices readily sliding past one another (for a general 
review, see Blatter {\it et al.}\,\cite{blatter}).   
Measurements of nonlocal effects probe the inherent length scales
of the problem and thus can be used to investigate issues such as
whether the decoupling and melting transitions occur simultaneously
\cite{fuchs}. 
The claims of Safar {\it et al.}\,\cite{safar} to have observed a 
sizable nonlocal effect in twinned ${\rm YBa_2Cu_3O_{7-\delta}}$ 
(YBCO) are based on two sets of measurements. 
In the first, which we refer to as the {\it top} geometry, a 
current is put into and drawn out of the top of a modified flux 
transformer while the voltage differences, $V_{top}$ and $V_{bot}$, 
are measured (see Fig.~\ref{fluxgeom1}). 
In the second, the {\it side} geometry, the current is withdrawn from 
the bottom and the voltages, $V_{left}$ and $V_{right}$, are 
measured.  
Safar {\it et al.}\,\cite{safar} find that the ratios $V_{bot}/V_{top}$ 
and $V_{right}/V_{left}$ both approach one as they near the melting 
transition.  
Taken individually either result might be explained by a local though 
anisotropic conductivity; but taken together the results are 
inconsistent with a local description. 
Safar {\it et al.}\,\cite{safar} confirm this by analyzing each data 
set as though the conductivity were local (the Montgomery analysis
\cite{montgomery}) and extracting from each the {\it apparent}
conductivity ratio $\sigma_{xx}^{(a)}/\sigma_{zz}^{(a)}$, finding
a huge discrepancy in these apparent ratios. 

Eltsev and Rapp \cite{eltsev} dispute the Safar {\it et al.} claim. 
They performed similar measurements but did not see $V_{right}/V_{left} 
\rightarrow 1$.
On the other hand, they may have seen nonlocal effects in a
{\it tilted} geometry in which the current is extracted from terminal
$6$ (instead of terminal $5$) and the ratio $(V_3-V_8)/(V_2-V_7)$
measured. 
In a comparison of twinned and untwinned YBCO, L\'opez 
{\it et al.}\cite{lopezall} find that in the untwinned YBCO, the 
strongest signature of nonlocality seen by Safar 
{\it et al.}\,\cite{safar}, $V_{bot} \approx V_{top}$, is no longer 
found for any significant temperature range above the ``melting"
transition. 
A feature of our studies below is that substantial nonlocal effects
are only  present when a characteristic length (presumably the
phase coherence length) is of the order of the sample thickness.
Now the phase coherence length scale along the field direction
(according to Ref.~\cite{pcm}) grows exponentially rapidly as the
temperature is lowered in such a way that the temperature interval
over which nonlocal effects might be visible in the vortex-liquid
region is only perhaps within $0.3K$ of the temperature at which
pinning drives the resistance rapidly to zero.
This is of the same order of magnitude as the rounding of the
zero-field transition due to sample inhomogeneities and as a
consequence it will be hard to disentangle the various effects
from each other in the untwinned results of
L\'opez {\it et al.}\cite{lopezall}.
If one supposes that the long length scale causing the nonlocality
in the twinned case is caused by a Bose-glass-like
mechanism~\cite{nelson1,wallin1}, then one would expect the coherence
length to increase only as a power law, and the width of the
temperature interval over which nonlocal effects are visible may
therefore be wider in the twinned case.

The assertions of nonlocal effects in ${\rm Bi_2 Sr_2 Ca Cu_2 O_x}$
(BSCCO) are less dramatic than those in YBCO. 
In their measurements on single-crystal BSCCO, Keener 
{\it et al.}\,\cite{keener} never observe the ratios $V_{bot}/V_{top}$ 
and $V_{right}/V_{left}$ simultaneously approaching one. 
Nevertheless, when they perform a Montgomery analysis on their data, 
they do see discrepancies in the apparent ratio $\sigma_{xx}^{(a)}/
\sigma_{zz}^{(a)}$. 
Conversely, measurements by Busch {\it et al.}\,\cite{busch} on 
single-crystal BSCCO and by Doyle {\it et al.}\,\cite{doyle} on BSCCO 
with columnar defects are claimed to be consistent with local 
resistivity. 
These seemingly contradictory results could be caused by approximations
used in the local analysis \cite{busch} and might be resolved by using
the full analysis or better suited approximations, such as the one
proposed by Levin \cite{levin}.   
A theoretical framework which could provide some quantitative 
analysis of these results---for example, by the extraction of a 
temperature-dependent length scale---would obviously be helpful in 
the interpretation of these and other results.
Our aim is to develop such a framework. 
We will approach the problem phenomenologically, attempting to relate 
the current-voltage characteristics to the form of the (nonlocal) 
conductivity. 

When a material has a nonlocal conductivity, the appropriate form of
Ohm's law is given by 
\begin{equation} j_\mu({\bf r}) = \int\,
\sigma_{\mu\nu}({\bf r , r'})\,{E_\nu({\bf r'})}\, d^3r',
\end{equation}
where symbols have their usual meaning in this context, and the 
integral is taken over the volume of the sample. 
For a nonlocal conductivity, $\sigma_{\mu\nu}({\bf r, r'}) \ne 
\sigma_{\mu \nu}\,\delta({\bf r-r'})$. 
In momentum space, for a translationally invariant system, this
relation becomes
\begin{equation}
\hat{\j}_\mu({\bf k}) = \hat{\sigma}_{\mu\nu}
({\bf k})\,\hat{E}_\nu({\bf k}),
\end{equation}
where $\hat\j_\mu({\bf k})$ is the Fourier transform of 
$j_\mu({\bf r})$, and similarly for the other quantities. 
It should be noted that nonlocal effects will only be observable
when the length scale of the nonlocality is of the same order
or larger than the distance between leads. 

The best known theoretical work on the subject is the
``hydrodynamic'' approach, expounded upon in general by Marchetti 
{\it et al.}\,\cite{marchetti} and applied specifically to the 
conductivity by Huse and Majumdar \cite{huse}. 
This theory is so-called because the nonlocal conductivity
$\hat\sigma_{\mu\nu}({\bf k})$ is expanded in a Taylor series in 
$\bf k$, and the expansion is terminated at order $k^2$; in
other words, the conductivity is taken to be of the form 
\begin{equation}
\label{hydro-form}
\hat\sigma_{\mu\nu}({\bf k}) = \hat\sigma_{\mu\nu}({\bf 0}) +
\eta_{\mu\beta\gamma\nu} \, k_\beta k_\gamma.
\end{equation}
We revert to the notation of Huse and Majumdar to facilitate
comparison with that work\,\cite{huse}; note that later 
works\,\cite{mou,blum,wortis} replace the $\eta$'s with $S$'s to 
prevent confusion with the viscosity tensor of the vortex-line 
liquid, which is related but distinct \cite{mou}. 
Unfortunately, this form for the conductivity is unphysical if
certain coefficients become negative, as shown by Blum and
Moore\cite{blum}. 
Huse and Majumdar always assume that the coefficients they use 
are positive. 
When and whether the coefficients are in fact positive or negative
will be discussed in more detail in Section~\ref{negative?}. 

The hydrodynamic analysis leads to a fourth-order partial differential
equation which reduces to Laplace's equation in the local limit
($\eta=0$).
It also supplies sufficient boundary conditions to solve for the
potential $V({\bf r})$.
Huse and Majumdar argue that there are discontinuities in the first
derivative of ${\bf E}({\bf r})$ at the surface.
When the conductivity (which in hydrodynamics is a differential
operator) is applied, the result is $\delta$-functions in the
current distribution at the surface, i.e. surface currents.
One then uses charge conservation, $\nabla \cdot {\bf j} =0$, to
translate this outcome into boundary conditions on $V({\bf r})$.
To handle the $\delta$-function it is convenient to integrate over
the surface as in the standard Gaussian pillbox
arguments \cite{jackson}---only here, because of the surface
current, the side surfaces of the pillbox contribute even as
the volume of the box is shrunk down to zero.
This gives what initially looks like an extra term in their 
boundary conditions.

Huse and Majumdar study a two-dimensional geometry modeling the 
flux transformer used in the experiments,  the $z$-axis of which
coincides with the $\hat{c}$-axis of the superconductor. 
They have performed a detailed analysis of the situation with one
non-zero viscosity coefficient, $\eta_{xzzx}$, which embodies the
interaction of pancake vortices moving in different $ab$ planes and
at different velocities. 
Somewhat surprisingly, their equation and boundary conditions are 
symmetric under $\eta_{xzzx} \leftrightarrow \eta_{zxxz}$ despite the 
fact that these coefficients would appear to represent very different 
physics. 

As an alterative to the hydrodynamic truncation of $\hat
\sigma_{\mu \nu}({\bf k})$, we consider an analysis based on Pad\'e
approximations to  $\hat \sigma_{\mu \nu}({\bf k})$.
It incorporates a more realistic large-${\bf k}$ behavior than the 
hydrodynamic approach and remains solvable. In principle, one can
approximate  $\hat \sigma_{\mu \nu}({\bf k})$ to any desired degree
of accuracy by using a sufficiently large-order Pad\'e approximation.
As the order of the approximation is increased, our technique of
solution continues to work, but the computing effort increases
rapidly.
One stage of the solution involves a partial differential
equation rather reminiscent of that occurring in hydrodynamics. 
In fact, in one instance we can recover the results of Huse and
Majumdar by means of a limiting procedure on the Pad\'e result. 
  
In the remainder of this paper we first discuss our motivation for 
wanting to improve upon and extend the work of Huse and Majumdar; 
this involves an examination of whether the relevant coefficients 
$\eta$ in the small-$k$ expansion of the conductivity are positive 
or negative. 
We investigate the current-voltage characteristics in a particular
geometry (the infinite-slab geometry), which allows us to comment on
whether positive or negative viscosity coefficients are needed to
explain experimental data like that of Safar {\it et al.}\,\cite{safar}
(Sec.~\ref{infinite}). 
The section following that contains details of work using Pad\'e
approximations to the conductivity. 
In Sec.~\ref{surfaces} we discuss the role of surfaces in determining 
the conductivity and how the analysis of Huse and Majumdar takes 
account of surfaces.
Appendices~\ref{inf-app} and \ref{pade-app} contain some 
calculational details, while in Appendix~\ref{bose} we outline
the Bose-glass scaling of the conductivities used in some of the
numerical work. 

\section{Are the viscosity coefficients negative or positive?} 
\label{negative?}

For stability, the conductivity tensor $\hat{\sigma}_{\mu\nu}(\bf k)$
must be a positive definite matrix. 
For this to be true of the hydrodynamic form, Eq.~(\ref{hydro-form}), 
certain of the viscosity coefficients $\eta_{\mu\alpha\beta\nu}$ 
must be positive; in particular, $\eta_{zzzz}$ and $\eta_{xxxx}$. 
The work of Mou {\it et al.}\,\cite{mou} and Blum and
Moore \cite{blum} shows that for high temperatures, these
coefficients are actually negative. 
Both works use the time-dependent Ginzburg-Landau equation as a 
starting point, so ``high temperatures'' in this context means near
the $H_{c2}(T)$  line. 
Thus to treat nonlocal conductivities in this region of the $H$-$T$
plane, one requires a model that can handle these so-called negative 
viscosities, ruling out the hydrodynamics approach. 
However, we expect substantial nonlocal behavior occurs only near the 
melting line, where some of the viscosities may very well be positive 
and hydrodynamics a viable approach. 

So what happens to $\hat \sigma_{\mu \nu}({\bf k})$ as the
temperature is lowered? 
The arguments of Mou {\it et al.}\,\cite{mou} suggest and the 
simulations of Wortis and Huse \cite{wortis} bear out that as the 
temperature is decreased, $\eta_{xxxx}$ changes sign, becoming  
positive. 
Imagine a plot of $\hat{\sigma}_{xx}(k_x)$ (see Wortis and 
Huse\cite{wortis}):  at high temperatures, $\hat{\sigma}_{xx}(k_x)$ 
increases monotonically as $k_x$ is decreased, but at low
temperatures, it develops a local maximum at a nonzero $k_x$ and
then falls to a finite value (the flux-flow value \cite{flux-flow})
at $k_x = 0$.

In the presence of a magnetic field, the $ab$-plane conductivity of
a type-$II$ superconductor does not diverge, because a current
causes the vortices to move, leading to dissipation, that is, a
non-zero resistance. 
The conductivity is thus enhanced if this movement of vortices is
impeded, for instance by pinning centers. 
The interaction of a vortex with other vortices may also inhibit its 
motion. 
However, for a uniform current the vortices all move together and
thus their mutual interactions play little role in hindering the  
center-of-mass motion. 
For a nonuniform current, on the contrary, the vortices are 
impelled to change their relative positions, and so their
interactions do inhibit this sort of motion. 
Therefore, the conductivity may be higher for a nonuniform current. 
These arguments suggest that at low temperatures where the
interactions become important the conductivity might be higher at
some nonzero $k_x$.  

The situation is different for conductivity along the $\hat{c}$-axis.
Like $\hat{\sigma}_{xx} (k_x)$, at high temperatures, 
$\hat{\sigma}_{zz}(k_z)$ increases monotonically as $k_z$ is lowered.
However, the current is now along the axis of the vortices, the
vortices are not forced to move, and this time the uniform
conductivity $\hat{\sigma}_{zz}({\bf k}={\bf 0})$ diverges as the
temperature is lowered.
Hence, there is no compelling reason to expect that a local maximum
will develop as temperature is lowered, and one might suspect that
$\eta_{zzzz}$ remains negative and therefore unsuited for the
hydrodynamic prescription at all temperatures.
This expectation for $\eta_{zzzz}$ seems to be borne out by
low-temperature calculations which have the Abrikosov lattice 
as a starting point \cite{sarah-eil} and also by the preliminary 
simulation results \cite{anne}. 

The proposition that some of the $\eta$'s may be negative at all
temperatures is one motive for wanting an alternative to
hydrodynamics; there are others. 
Hydrodynamics is a simple approximation to the actual conductivities, 
but we do not know how good an approximation it is. 
Plus, there is no obvious way to improve upon it---one might 
include terms of order $k^4$, but this necessitates additional
boundary conditions to specify the solution, and it is unclear
what they would be. 
In the Fourier representation of a function, the small-$k$ terms
model well its bulk properties, but higher-$k$ terms are needed
to capture the behavior at the boundaries.
So one might think a procedure focusing on small-$k$ would do well
in the bulk and perhaps less satisfactorily at the surface. 
But hydrodynamics involves a differential equation, and its solution,
even in the bulk, is determined by boundary conditions, i.e.
the surface.  
Thus it is crucial to treat the surface properly---even more
so, since in the experiments at issue here, all of the measurements
are taken at the surface.
Huse and Majumdar do take some account of surfaces, and in doing so 
find surface currents, but their theory is unable to make any 
prediction about the length scale over which these currents might 
flow. 
Such restrictions as these provide the impetus to go beyond 
hydrodynamics. 

There is one further comment that is useful to make before
proceeding to some concrete calculations. 
While we require that $\hat{\sigma}_{\mu\nu}(\bf k)$ be a positive 
definite matrix, this does not imply that $\sigma_{\mu\nu}(\bf r)$ 
must always be positive. 
Although conductivities taking on negative values seems a little odd
at first sight, the simulations of Wortis and Huse\cite{wortis} find
that $\sigma_{xx}(x,k_y=0)$ can be negative over a range of a few
inter-vortex spacings. 
In fact, {\it any} nonlocality in the conductivity implies that
either the real-space conductivity or resistivity (or both) take
on negative values at some points. 
By definition, the conductivity matrix is the inverse of the
resistivity matrix, which implies that in Fourier space, 
$\hat{\sigma}_{\mu\nu}(\bf k) \hat{\rho}_{\nu\alpha}(\bf k) = 
\delta_{\mu\alpha}$ ($\delta_{\mu\alpha}$ is the Kronecker delta 
function and the summation convention is used). 
For example, if $\sigma_{xy}=\sigma_{xz}=0$ we have in real space  
\begin{equation}
\label{real-space}
\int d^3r'\,\sigma_{xx}({\bf r}-{\bf r'})\,
\rho_{xx}({\bf r'}-{\bf r''}) =
\delta(\bf r-r'').
\end{equation}
If ${\bf r} \neq {\bf r''}$, the right hand side of
Eq.~(\ref{real-space}) is zero; for the left hand side to be
zero, some cancellation is needed.   
However, there would be no cancellation if $\sigma_{xx}
({\bf r}-{\bf r'})$ and $\rho_{xx}({\bf r'}-{\bf r''})$ are both 
everywhere positive. 
Thus, one or both has negative regions. 
For experimental setups in which the current is distributed 
throughout the sample, this feature may not manifest itself in 
the voltage distribution; thus, Wortis and Huse\,\cite{wortis} 
have proposed geometries with very localized currents in order 
to look for it.            

\section{The infinite-slab geometry and positive viscosities}
\label{infinite}

\begin{figure}[tbp]
\centerline{\epsfxsize=7.0cm 
\epsfbox{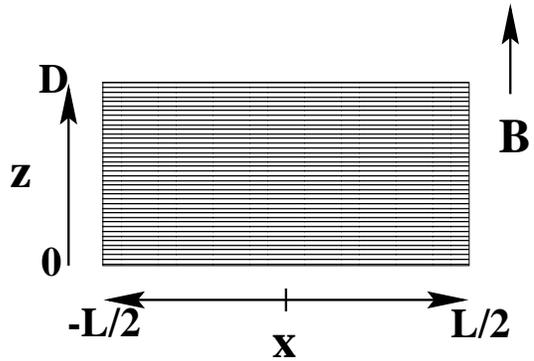}}
\vspace{1.0cm}
\caption{\label{geom1}The geometry used for our calculations of
the potential $V(x,z)$. 
The $\hat{c}$-axis of the superconductor is aligned along the
$z$-axis (called the $y$-axis in the notation of Huse and Majumdar).}
\end{figure}

The aim of this section is to show that for the choice of
conductivity given below, it is necessary to have {\it positive}
viscosity coefficients to produce the strongly nonlocal behavior
seen by Safar {\it et al.}\,\cite{safar}: $V_{bot}/V_{top}
\rightarrow 1$ and $V_{right}/V_{left} \rightarrow 1$
simultaneously.
Let us consider the two-dimensional geometry shown in
Fig.~\ref{geom1}, with the lateral dimension $L \rightarrow
\infty$; taking this limit eliminates one set of boundary effects.  
Furthermore, let us postulate conductivities of the form
\begin{mathletters}
\label{cond1}
\begin{eqnarray}
\hat{\sigma}_{xx}(\bf k) &=& \Sigma_x ; \\
\hat{\sigma}_{zz}(\bf k) &=& \hat{\sigma}_{zz}(k_x),
\end{eqnarray}
\end{mathletters}
that is, $\hat \sigma_{xx}$ is a constant (i.e. local) and $\hat 
\sigma_{zz}$ is a function of $k_x$ alone. 
The function $\hat \sigma_{zz}(k_x)$ may include a constant piece;
moreover, that constant, as well as $\Sigma_x$, may represent both
superconducting and normal contributions to the local
conductivity. 
Restricting the wavevector dependence to $k_x$ alone enables us to
solve for the potential $V(x,z)$ via Fourier transformation.   

To determine $V(x,z)$ for a given input current, first relate
$V(x,z)$ to the components of the current 
\begin{mathletters}
\begin{eqnarray}
\label{inf-curr}
j_x(x,z) &=& - \Sigma_x \, \partial_x V(x,z); \\ 
\label{inf-currb}
j_z(x,z) &=&  - \int_{-\infty}^\infty \hspace{-0.2cm}
\sigma_{zz}(x-x') \, \partial_z V(x',z) \, dx',
\end{eqnarray}
\end{mathletters}
using $E_{\mu}=-\partial_{\mu} V $. 
Next note that in the steady state, the continuity equation is 
${\bbox \nabla} \cdot {\bf j} = 0$, which in this case is  
\begin{equation}
\label{inf-cont}
\Sigma_x \, \partial_x^2 V(x,z) 
+\int_{-\infty}^\infty \hspace{-0.3cm}\sigma_{zz}(x-x')\,
\partial_z^2 V(x',z) \, dx' = 0.
\end{equation}
Fourier transforming Eq.~(\ref{inf-cont}) with respect to $x$
yields 
\begin{equation}
\label{inf-ft}
- \Sigma_x \, k_x^2 \,\hat V(k_x,z) +
\hat \sigma_{zz}(k_x) \,\partial_z^2 \hat V(k_x,z) =0, 
\end{equation} 
where we have used the fact that the transform of the convolution
is the product of the transforms.  
The solution of this differential equation is  
\begin{equation}
\label{inf-sol}
\hat V(k_x,z) = A(k_x) \cosh [\kappa(k_x) z] + 
B(k_x) \sinh [\kappa(k_x) z], 
\end{equation}
where 
\begin{equation}
\kappa^2(k_x) = \frac{ \Sigma_x \, k_x^2 }{\hat \sigma_{zz}(k_x)}.
\end{equation}
One determines the functions $A(k_x)$ and $B(k_x)$ from the boundary 
conditions; toward this end, it is convenient to Fourier transform
the expression for $j_z$ (Eq.~(\ref{inf-currb}))
\begin{equation}
\label{inf-bc}
\hat\sigma_{zz}(k_x) \, \partial_z \hat V(k_x,z) = - \hat j_z(k_x,z). 
\end{equation} 

Imposing the boundary conditions appropriate for the top geometry, 
namely $j_z(x,0)= 0$ and $j_z(x,D) = J_{T}(x)$, finding
$\hat V(k_x,z)$ and taking the inverse transform yields  
\begin{equation}
\label{vtop}
V_{T}(x,z) = -\int^\infty_{-\infty} \frac{dk_x}{2\pi} 
\frac{\hat{J}_{T}(k_x) \,{\cosh}(\kappa z)
\,{\rm e}^{i k_x x}}
{\kappa \, \hat\sigma_{zz}(k_x) \, \sinh (\kappa D)}.
\end{equation}
Note that charge conservation implies $\hat{J}_{T}(0) = 0$,
and hence the integral above converges at $k_x = 0$. 
For the side geometry, similar manipulations using the boundary 
conditions $j_z(x,0) = j_z(x,D) = J_{S}(x)$ give 
\begin{equation}
\label{vside}
V_{S}(x,z)= \int^\infty_{-\infty}\frac{dk_x}{2\pi}
\frac{\hat J_S (k_x) 
\sinh \left[\kappa \left( \frac{D}{2}-z \right) \right]
{\rm e}^{ik_x x}}
{\kappa \, \hat \sigma_{zz}(k_x) \, \cosh
\left( \frac{\kappa D}{2} \right)} .
\end{equation}

Analytically the $x$-axis decay length is controlled by the pole
structure of the above integrals (see Appendix~\ref{inf-app} for
more details).  
Rather generically, this length grows if the viscosity coefficient
(the coefficient of $k_x^2$ in the small-$k_x$ expansion of
$\sigma_{zz}(k_x)$) is positive, leading to features such as
$V_{right}/V_{left} \rightarrow 1$. 
Increasing this coefficient ($\eta$) also changes the ratio
$V_{bot}/V_{top}$ though not necessarily in a monotonic fashion.  
However, there is another way to ensure that $V_{bot}/V_{top}
\rightarrow 1$: this is simply to make $\sigma_{zz}$ very large,
which applies even in the local limit.  
Thus, to obtain results like those of Safar
{\it et al.}\,\cite{safar}, we expect that a $\hat \sigma_{zz}(k_x)$
which grows large and has a positive viscosity coefficient at low 
temperatures is required. 

\begin{figure}[tbp]
\centerline{\epsfxsize=7.0cm 
\epsfbox{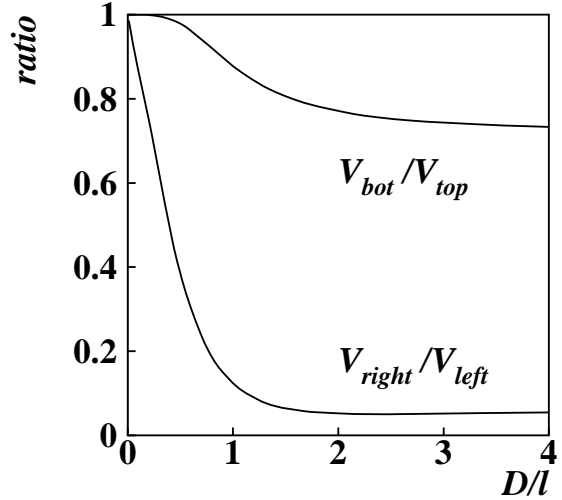}}
\vspace{1.0cm}
\caption{\label{cond}The ratios of $V_{bot}/V_{top}$ and
$V_{right}/V_{left}$ for the conductivity with positive
viscosity coefficients, the form of which is provided in the
text. The boundary conditions used were $J_{T}(x) = J_0 \left[
\delta(x - 2 L_c) - \delta(x + 2 L_c) \right]$ and $J_{S}(x) = -
J_0\,\delta(x + 2 L_c)$ ($ 4 L_c $ is defined to be the distance
between the current inputs). 
The voltages $V_{top}$ and $V_{bot}$ were measured at $x = \pm L_c$
as were the voltages $V_2$, $V_3$, $V_6$ and $V_7$.
The parameters used are $\Sigma_x = 5$, $\sigma_z^{(n)}= 1$,
$C = 1$, $D = 1$ and $ L_c = 1 $; this corresponds to measuring
lengths in terms of the thickness of the sample and conductivities
in terms of  $\sigma_z^{(n)}$, as will also be done in all the
other figures. }
\end{figure}

These arguments have been checked numerically for a variety of 
conductivities. 
One example is shown in Fig.~\ref{cond}. 
We have put all of the temperature dependence of the conductivity
into the length scale $\ell$ which is presumed to increase as
temperature decreases and choose a conductivity of the form
$\hat{\sigma}_{zz}(k_x) = \sigma_z^{(n)}+ C\ell^2(1+2k_x^2 \ell^2)
/(1 + k_x^2 \ell^2)$, which meets the above criterion as $\ell$
increases. 
Note that we have also included $\sigma_z^{(n)}$, a local term that 
does {\it not} scale with $\ell$. 
Fig.~\ref{cond} shows a plot of $V_{bot}/V_{top}$ and $V_{right}/
V_{left}$ as a function of $1/\ell$, for this choice. 
It can be seen that as $\ell \rightarrow \infty$, the two ratios
do indeed approach one.  

\begin{figure}[tbp]
\centerline{\epsfxsize=7.0cm 
\epsfbox{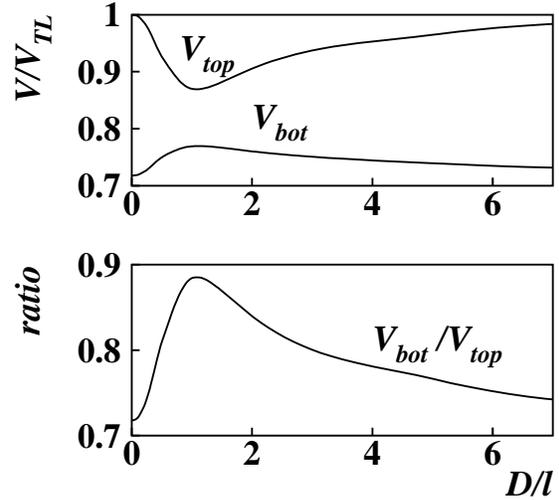}}
\vspace{1.0cm}
\caption{\label{nonsense}The voltages $V_{top}$ and $V_{bot}$ and
their ratio for the conductivity $\hat{\sigma}_{zz}(k_x) =
\sigma_z^{(n)}+C\ell {\rm e}^{-k_x^2 \ell^2}$, with the same values
of parameters as in Fig.~\protect\ref{cond}. By definition,
$V_{TL}$ is defined to be $V_{top}$ in the local case for this
geometry, which can be calculated from Eq.~(\protect\ref{localcase}).
Note that $V_{bot}/V_{top}$ does not tend to $1$ as $\ell
\rightarrow \infty$.}
\end{figure}

\begin{figure}[tbp]
\centerline{\epsfxsize=7.0cm 
\epsfbox{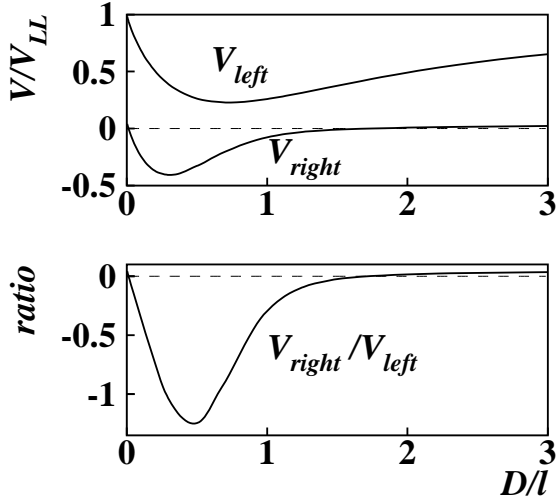}}
\vspace{1.0cm}
\caption{\label{nonsense1}The voltages $V_{left}$ and $V_{right}$
and their ratio for the conductivity $\hat{\sigma}_{zz}(k_x) = 
\sigma_z^{(n)}+ C\ell {\rm e}^{- k_x^2 \ell^2}$, with the same
parameters as for Figs.~\protect\ref{cond} and
\protect\ref{nonsense}.
Here, $V_{LL} $ is defined to be $V_{left}$ for the local case,
Eq.~(\protect\ref{localcase}). 
The notable feature here is that $V_{right}$ becomes negative. }
\end{figure}

Using conductivities with negative viscosity gives results such as 
those shown in Figs.~\ref{nonsense} and \ref{nonsense1}. 
In this particular case, the conductivity used was
$\hat{\sigma}_{zz}(k_x) = \sigma_z^{(n)}+C\ell
{\rm e}^{- k_x^2 \ell^2}$; however, the results are typical of
conductivities with negative viscosity. 
Finding negative voltages is not necessarily unphysical
(see the discussion surrounding Eq.~(\ref{real-space})) but the
results clearly do not give us the strongly nonlocal behavior seen
by Safar {\it et al.}\,\cite{safar}. 

Now let us consider the difference between the experimental data
taken from twinned YBCO \cite{safar} and untwinned YBCO
\cite{lopezall}. 
Recall that for the twinned YBCO, the $V_{top}$ and $V_{bot}$ curves
meet at some temperature $T_{th}$, and they continue toward zero 
together as the temperature is lowered. 
For the untwinned YBCO, on the other hand, the curves only meet 
just before dropping sharply to zero.  
We can reproduce some of these features by choosing the scaling
forms for the conductivity appropriately, depending upon whether
the sample is twinned or untwinned. 
For the twinned case, we use the Bose-glass scaling forms, which are 
discussed in Appendix~\ref{bose}. 
Although these are supposed to be valid in the presence of columnar
defects, and not strictly twin planes, we use them here since there
is currently no better alternative, and there is at least some
experimental evidence\cite{evidence} to support a Bose-glass
transition in twinned YBCO. 
For the untwinned case, we will use the same forms used to
generate Fig.~\ref{cond}.

We choose conductivities given by
\begin{equation}
\label{scal}
\begin{array}{ll}
\mbox{Untwinned:} & \left\{
\begin{array}{l}
\hat{\sigma}_{xx} = \sigma_x^{(n)} \\ 
\hat{\sigma}_{zz} = \sigma_z^{(n)} +
\displaystyle{\frac{C_1 \ell^2 \left(1+2 k_x^2 \ell^2
\right) }{1 + k_x^2 \ell^2}}, 
\end{array} \right. \\ \\
\mbox{Twinned:} & \left\{
\begin{array}{l}
\hat{\sigma}_{xx} = \sigma_x^{(n)} + C_2 \ell^4\\ 
\hat{\sigma}_{zz} = \sigma_z^{(n)}  + 
\displaystyle{\frac{ C_1 \ell^6 \left(1+2k_x^2
\ell^2 \right)}{1 +  k_x^2 \ell^2}}.
\end{array} \right.
\end{array}
\end{equation}
Both choices have the same overall form with positive viscosity
coefficients, the only difference is in how the constants scale
with $\ell$. 

\begin{figure}[tbp]
\centerline{\epsfxsize=7.0cm 
\epsfbox{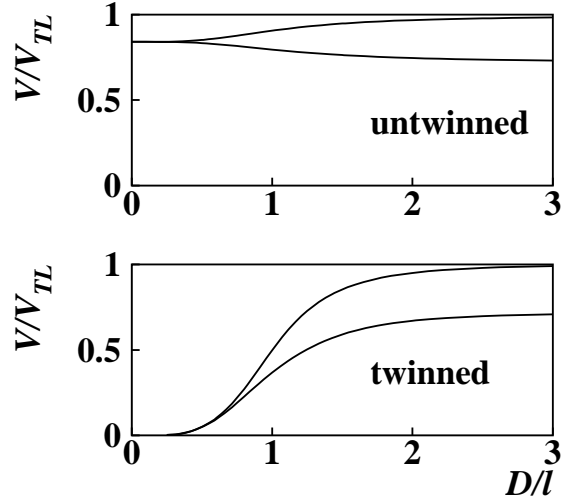}}
\vspace{0.5cm}
\caption{\label{scaled}The results for $V_{top}$ and $V_{bot}$ as a
function of $\ell$, using the conductivities of
Eq.~(\protect{\ref{scal}}) with parameters $\sigma_x^{(n)} = 5$,
$\sigma_z^{(n)} = 1$, $ D = 1$, $C_1 = 1$ and $ C_2 = 5$. The upper
curve in each plot is $V_{top}$, and the lower is $V_{bot}$.}
\end{figure}

The results obtained are shown in Fig.~\ref{scaled}. 
Firstly, it can be seen that in both instances, $V_{top}/V_{bot} 
\rightarrow 1$ as the length scale $\ell$ is increased. 
However, for the twinned case, both $V_{top}$ and $V_{bot}$ tend
to zero together, whereas in the untwinned case, they tend to some 
finite value. 
If one identifies the sharp drop in the corresponding experimental 
results as the transition from a vortex liquid to solid, then this 
is the correct behavior. 
The voltages in Fig.~\ref{scaled} level off as $\ell$ grows small,
whereas the voltages measured in the experiments continue to
rise as the temperature is increased.
This discrepancy is due to our neglect of, among other things,
the temperature dependence of the normal component, which can
be nontrivial~\cite{normal}.
Nevertheless, the results show that varying the scaling
behavior can account for some of the differences between the 
twinned and untwinned YBCO. 

\section{Work with Pad\'e approximations} \label{pade}  

In this section, we lift some of the restrictions in the previous
section by considering a flux-transformer geometry with finite
lateral dimension $L$ and by generalizing the nonlocal
conductivities of Eq.~(\ref{cond1}) to allow either
$\hat\sigma_{xx}$ or $\hat\sigma_{zz}$ to be nonlocal and depend
on either $k_x$ or $k_z$.
As opposed to the infinite-slab geometry, we could not make 
progress with general conductivities, thus we elect to use Pad\'e
approximations to the conductivities. 
Two important advantages to using Pad\'e forms are: 1)
they capture the high-${\bf k}$ behavior as well as the
low-${\bf k}$ behavior of the true conductivity and 2) the resulting
equations are analytically soluble. 
In addition, the corresponding real-space conductivities can be
chosen to be decaying exponentials, which, according to Wortis and
Huse\cite{wortis}, could be the correct form for the
high-temperature regime.

We will consider four different cases: 
\begin{mathletters}
\label{cases}
\begin{eqnarray}
\label{c1}
1.\,\, \hat{\sigma}_{xx}({\bf k}) = \Sigma_x, \ \ \ 
&&\hat{\sigma}_{zz}({\bf k}) = \Sigma_z + \frac{\Delta_z^z}
{1+k_z^2 \ell^2} ; \\
\label{c2}
2.\,\, \hat{\sigma}_{zz}({\bf k}) = \Sigma_z, \ \ \ 
&&\hat{\sigma}_{xx}({\bf k}) = \Sigma_x +\frac{\Delta_x^z}
{1+k_z^2 \ell^2} ; \\
\label{c3}
3.\,\, \hat{\sigma}_{zz}({\bf k}) = \Sigma_z, \ \ \ 
&&\hat{\sigma}_{xx}({\bf k}) = \Sigma_x +\frac{\Delta_x^x}
{1+ k_x^2 \ell^2}; \\
\label{c4}
4.\,\, \hat{\sigma}_{xx}({\bf k}) = \Sigma_x, \ \ \ 
&&\hat{\sigma}_{zz}({\bf k}) = \Sigma_z + \frac{\Delta_z^x}
{1+ k_x^2 \ell^2}. 
\end{eqnarray}
\end{mathletters}
Note that the momentum dependence is upon $k_z$ in Cases $1$ and
$2$ whereas it is upon $k_x$ in Cases $3$ and $4$ and that
$\hat{\sigma}_{xx}$ is local in Cases $1$ and $4$ while
$\hat{\sigma}_{zz}$ is local is Cases $2$ and $3$.
In the Pad\'e form, $\Sigma$ is the conductivity as $k
\rightarrow \infty$, $\Sigma+\Delta$ is that at $k=0$, and $\ell$
is a length scale. 
We set out to solve ${\bbox \nabla} \cdot {\bf j} = 0$ subject to 
the usual boundary conditions on the current at the surface. 
Each of these problems involves an integro-differential equation,
which can be converted into a partial differential equation with
linear coefficients, which can in turn be solved by separation of
variables.
This procedure is outlined in Appendix~\ref{pade-app}. 
Below we point out some of the distinguishing features of
the various cases.

{\bf Case $1$.} 
With conductivities of the form in Eq.~(\ref{c1}), solving 
${\bbox \nabla} \cdot {\bf j} = 0$ leads to the partial
differential equation 
\begin{equation}
\label{fourth-1}
\left[\partial_z^4 + \frac{\Sigma_x}{\Sigma_z}
\partial_z^2 \partial_x^2 
-\frac{\left(1+\gamma_z^z \right)}{\ell^2} \partial_z^2   
-\frac{\Sigma_x}{\Sigma_z \ell^2} \partial_x^2  \right] V=0,
\end{equation}
where $\gamma_{\mu}^{\nu}$ is a dimensionless variable given
by $\gamma_{\mu}^{\nu} =\Delta_{\mu}^{\nu}/\Sigma_{\mu}$.
Separation of variables, $V(x,z) = X(x) \, Z(z)$, then 
yields
\begin{mathletters}
\begin{eqnarray}
\label{xeq-text}
&& \left( \frac{d^2}{dx^2}  + k^2 \right) X =0, \\
\label{zeq-text}
&& \left\{ \frac{d^4}{dz^4}  - 
\left[\frac{\Sigma_x k^2}{\Sigma_z}  +
\frac{ \left(1 + \gamma_z^z \right) }{\ell^2}\right]
\frac{d^2}{dz^2}  
+ \frac{ \Sigma_x k^2}{\Sigma_z \ell^2 } \right\} Z =0. 
\end{eqnarray}
\end{mathletters}
The solution of Eq.~(\ref{xeq-text}) is $X(x)=A\cos(kx)+B\sin(kx)$.
Applying the boundary condition that no current enters on the sides,
i.e. $\partial_x V(\pm L/2,z) = 0$, gives $k=n \pi/L$ with $B=0$
for even $n$ and $A=0$ for odd $n$.
When $n > 0$, the solution of Eq.~(\ref{zeq-text}) is
\begin{equation}
\label{formofsoln-text}
Z(z) = 
A {\rm e}^{z/\xi_+} + B{\rm e}^{-z/\xi_+} +
C{\rm e}^{z/\xi_-} + E{\rm e}^{-z/\xi_-} ,
\end{equation}
which has two length scales given by  
\begin{eqnarray}
\xi^{-2}_{\pm}(n) &=& \frac{1}{2}  \left\{  
\frac{ n^2}{ \lambda_z^2} + \frac{1+\gamma_z^z}
{\ell^2}  \right. \nonumber \\
&& \left. \pm \left[\left(\frac{n^2}{\lambda_z^2} +  
\frac{1+\gamma_z^z}{\ell^2} \right)^2 -
\frac{4  n^2 }
{ \ell^2 \lambda_z^2} \right]^{1/2} \right\},
\end{eqnarray}
where $\lambda_z=\sqrt{\Sigma_zL^2/\pi^2\Sigma_x}$ is a length
scale occurring in the local limit ($\gamma_z^z=0$). 
 Note that if $\gamma_z^z >0$ (corresponding to a negative
viscosity coefficient) the $\xi$'s are real; whereas if
$\gamma_z^z <0$ (corresponding to a positive viscosity
coefficient) the $\xi$'s can become complex.

{\bf Case $2$.} 
For conductivities given by Eq.~(\ref{c2}), the corresponding
differential equation is 
\begin{equation}
\label{fourth-2}
\left[\partial_z^4 
+ \frac{\Sigma_x}{\Sigma_z} \partial_z^2 \partial_x^2 
-\frac{1}{\ell^2} \partial_z^2   
-\frac{\Sigma_x}{\Sigma_z}
\frac{\left(1  + \gamma_x^z \right)}{\ell^2}
\partial_x^2 \right]V=0.
\end{equation}
Proceeding with separation of variables, the functions $X_n(x)$ 
are exactly the same as in Case $1$, and the functions $Z_n(z)$
have the same form as in Eq.~(\ref{formofsoln-text}), but the
two length scales $\xi_{\pm}$ are now given by
\begin{equation}
\label{omega2}
\xi^{-2}_{\pm} = \frac{1}{2}  \left\{  
\frac{ n^2 }{\lambda_z^2} + \frac{1}{\ell^2}  
\pm \left[\left(\frac{n^2}{\lambda_z^2} - 
\frac{1}{\ell^2} \right)^2 - 
\frac{4  n^2 \gamma_x^z  }
{ \ell^2 \lambda_z^2}\right]^{1/2} \right\}.
\end{equation}
As opposed to Case $1$, this time when $\gamma_x^z< 0$ 
the $\xi$'s are real; and when $\gamma_x^z > 0$ the $\xi$'s 
may become complex.  

{\bf Case $3$.} Turning the example with conductivities given
by Eq.~(\ref{c3}) into a partial differential equation yields
\begin{equation}
\label{fourth-3}
\left[\partial_x^4  + 
\frac{\Sigma_z}{\Sigma_x} \partial_x^2 \partial_z^2  -
\frac{\left(1 + \gamma_x^x \right)}{\ell^2} 
\partial_x^2 
- \frac{\Sigma_z}{\Sigma_x \ell^2} \partial_z^2 \right]V = 0 .
\end{equation}
And separation of variables leads to 
\begin{mathletters}
\begin{eqnarray}
&& \left\{ \frac{d^4}{dx^4}  +
\left[ \frac{\Sigma_z \kappa^2}{\Sigma_x} - 
\frac{\left(1 + \gamma_x^x \right)}{\ell^2} \right]
\frac{d^2}{dx^2}
- \frac{\Sigma_z \kappa^2 }{\Sigma_x \ell^2} \right\} X = 0; \\
&& \left( \frac{d^2}{dz^2}  - \kappa^2 \right) Z = 0.
\end{eqnarray}
\end{mathletters}
We see here that this situation differs from the previous two in
that the solution of $X(x)$ is no longer simply sines and cosines,
and similarly $\kappa$ has become nontrivial. 
 This feature complicates the application of the boundary
conditions and the summing over eigenfunctions necessary to
achieve a complete solution. 

{\bf Case $4$.} Conductivities of the form Eq.~(\ref{c4}) lead to 
\begin{equation}
\label{fourth-4}
\left[ \partial_x^4  + 
\frac{\Sigma_z}{\Sigma_x} \partial_x^2 \partial_z^2  -
\frac{1}{\ell^2}  \partial_x^2 
- \frac{\Sigma_z}{\Sigma_x}
\frac{\left(1 +\gamma_z^x \right)}{\ell^2} 
\partial_z^2 \right]V = 0.   
\end{equation}
The method of solution is like that for Case $3$, including the
nontrivial values of $\kappa$.

{\bf Results.}
Let us now consider the results of using Pad\'e
approximants. 
We look at the behavior of $V_{top}$ and $V_{bot}$ as a function of 
$\ell$, as in the previous section. 
In Cases $1$ and $2$, we input a current
\begin{equation}
\label{it}
I_{top}(x) = I_1 \sin(\pi x /L); \ \ \ \ 
I_{bot}(x)=0
\end{equation}
for the top geometry, and
\begin{equation}
\label{is}
I_{top}(x) = I_{bot}(x) = I_0 \left[1 + \sin(\pi x/L)\right]
\end{equation}
for the side geometry. 
These currents are chosen because they roughly approximate the
experimental inputs but involve the minimum number of Fourier
components, which simplifies the calculation.  
If we input the same currents in Cases $3$ and $4$, our solution
will involve an infinite number of terms since our formula for
$V(x, z)$ is not in the form of a Fourier series but is a sum over
more complicated eigenfunctions. 
We instead expand the current in terms of these eigenfunctions.  
In fact, we once again choose the input currents that minimize the 
number of terms in the sum over eigenfunctions. 
In the top geometry, the current is chosen to be proportional to 
the eigenfunction corresponding to $\kappa_1$, where $\kappa_1$ is 
the eigenvalue which tends to $\pi /L$ in the local limit. 
For the side geometry, we choose the combination of two
eigenfunctions that tends to the current of Eq.~(\ref{is}). 
For Case 3, this unfortunately means that we vary the input current
as we vary $ \ell$; at $ \ell = 0 $, the input current is the same
as for Cases 1 and 2, but this smoothly evolves so that in the
limit of $\ell \rightarrow \infty$, the input current is
$I_{top} \approx I_2 \sin( 2 \pi x/L)$.
The side geometry is affected similarly. 
Hence, we must bear in mind that the input current changes 
significantly in Case $3$. 
In Case 4, on the contrary, the variation of input current with
$ \ell $ does not appear to be as substantial.
The voltages were calculated at $z = 0$, $z = D$ and 
$x = \pm 0.3L$. 
For each case, we did two sets of calculations: one using the
conductivities as written in Eqs.~(\ref{cases}) with $ \Delta $
scaling as $ \ell^2 $, and the second using the scaling forms
appropriate for a Bose glass; as given in Appendix~\ref{bose} by
Eqs.~(\ref{case1sca})---(\ref{case4sca}).

The first point to make about all of the results is that 
we never found $V_{right}/V_{left} \rightarrow 1$ so long as 
$\Delta > 0$, which corresponds to a negative viscosity coefficient. 
Hence none of these forms give us the strongly nonlocal 
behavior seen by Safar {\it et al.}\,\cite{safar} (though
perhaps they do resemble other results \cite{eltsev,keener}).  
We showed in the infinite-slab case (Sec. III), in which the
conductivities corresponded most closely to Case 4, that we could
only model Safar's results with positive viscosity coefficients.
The results of this section suggest that this statement may hold
for all of the cases.

\begin{figure}[tbp]
\centerline{\epsfxsize=7.0cm 
\epsfbox{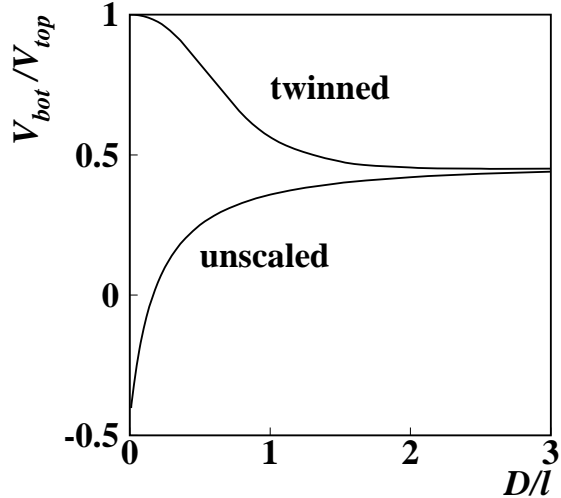}}
\vspace{0.5cm}
\caption{\label{negative} The results obtained for the ratio
$V_{bot}/V_{top}$ for Case $2$.
The conductivities used are given by Eq.~(\protect{\ref{c2}}) for
the lower curve, and Eq.~(\protect{\ref{case2sca}}) for the
upper curve (which has the appropriate scaling for a twinned sample).
The parameters used are $\Sigma_x = \sigma_x^{(n)} = 5$,
$\Sigma_z = \sigma_z^{(n)} = 1$, $ \ell_\parallel \equiv \ell $,
$ D = 1 $, $ L = 5 $, $C_1 = 1$, $C_2 = 5 $ and $ \Delta_x^z = 5
\ell^2$ --- note the scaling with $ \ell$ which is present.
For the twinned case, $V_{bot}/V_{top}$ does in fact tend to $1$
as $\ell \rightarrow \infty$. }
\end{figure} 

Secondly, we look at the difference between the Bose-glass scaled
and unscaled results.
We have already seen one example of this comparison in
Fig.~\ref{scaled}.
Another appears in Fig.~\ref{negative}, this time for Case $2$.
The curves are remarkably different considering they came from the
same overall form of the conductivity, Eqs.~(\ref{c2}), which shows
how vastly different behaviors can be modeled by the same form,
making it difficult to extract detailed information on the
conductivity from the experimental data.

Recall that the analysis of Huse and Majumdar~\cite{huse},
discussed in the introduction, has a symmetry under
$\eta_{xzzx} \leftrightarrow \eta_{zxxz}$.
If this symmetry applies in the Pad\'e analysis, it would
correspond to a symmetry between Cases $2$ and $4$. 
We have looked for such a symmetry.
The comparison is complicated by the fact that the eigenfunctions
$X_n(x)$ are different in the two cases, as mentioned above. 
However, using conductivities given by Eq.~(\ref{cases}) led
to qualitatively different features for Cases $2$ and $4$---while
admittedly the Case $4$ input currents change in this analysis,
we do not expect the difference to affect the general features of
the results. 
Thus, the Pad\'e forms do not seem to share the symmetry found 
in hydrodynamics. 
Furthermore, this lack of symmetry seems to persist even when 
we choose Pad\'e forms with positive viscosity coefficients 
that should correspond more closely to the hydrodynamic case.   

Notice that in the unscaled data in Fig.~\ref{negative} $V_{bot}$
goes negative.
Although the potential reverses sign, calculations reveal that
the current flow is always from left to right, even at the bottom
of the sample.
In Sec.~\ref{negative?} we showed that if the conductivity is
nonlocal, then either the real-space resistivity or conductivity
is negative at some point.
We believe that the sign reversal in the unscaled data is simply
a consequence of this fact.
It should be noted that voltage reversals have in fact been
measured by S. Aukkaravittayapun {\it et al.}\,\cite{reversals};
however, it seems unlikely that nonlocal conductivity accounts for
their results.

Before concluding this section, we note that it is possible to extend
the Pad\'e forms considered above to higher levels of approximation.
Suppose one simply adds a second Pad\'e term to the first one, 
for instance, 
\begin{equation}
\hat{\sigma}_{xx}(k_x) = \Sigma_x + \frac{\Delta_1}
{1 + k_x^2 \ell_1^2} + \frac{\Delta_2}{1 + k_x^2 \ell_2^2}.
\end{equation}
There are now two length scales, and one can choose the parameters 
so that the conductivity has a positive viscosity and maintain 
the property $\sigma_{xx}(0) > \sigma_{xx}(k_x \rightarrow \infty)$.  
It is still possible to solve the equation ${\bbox \nabla} \cdot 
{\bf j} = 0$ in a manner similar to the examples discussed above. 
However, one would need to differentiate the integro-differential 
equation four times instead of twice in order to eliminate the 
two integral terms.  
As a result, one ends up with a sixth-order equation and a lot 
more algebra. 
We have not pursued this avenue.

\section{Surface Considerations}
\label{surfaces}

We begin this section on surface considerations with a comparison 
of the voltage and current distributions found using the hydrodynamic 
approach with those found using the Pad\'e approach. 
Let us consider Case $2$ from the previous section making the
following parameter choices
\begin{mathletters}
\label{small-l}
\begin{eqnarray}
\Sigma_x = \sigma_x^0 + \eta \ell^{-2}, \\
\Delta_x^z = - \eta \ell^{-2}, 
\end{eqnarray}
\end{mathletters}
so that the small-${\bf k}$ expansion of the conductivity is 
\begin{equation}
\sigma_{xx}({\bf k}) = \sigma_x^0 + \eta k_z^2 -
\eta \ell^2 k_z^4 + O(\ell^4 k_z^6). 
\end{equation}
Note that the $\ell \rightarrow 0$ limit coincides with the
truncation used in the hydrodynamic approach. 
The fourth-order equation derived for Case $2$, Eq.~(\ref{fourth-2}), 
becomes 
\begin{equation}
\label{case2-small-l}
\left[\ell^2 \Sigma_z \partial_z^4 + 
\left(\eta +\ell^2 \sigma_x^0 \right) \partial_z^2 \partial_x^2  -
\sigma_x^0 \partial_x^2  - \Sigma_z \partial_z^2 
 \right]V = 0 .
\end{equation}
If the terms proportional to $\ell^2$ are dropped, the equation 
is identical to the one studied by Huse and Majumdar\,\cite{huse,bc}. 
However, notice the small parameter, $\ell^2$, multiplies the 
highest derivative, $\partial_z^4 V$.  
This is the classic scenario for the development of a {\it boundary 
layer}, a small region in which $V$ varies quite rapidly and 
in which $\ell^2 \partial_z^4 V$ is not negligibly small.\cite{bender}
Recall that our analysis of Case $2$ produced two length scales
given by Eq.~(\ref{omega2}). 
In the small-$\ell$ limit, $\xi_+$ is a small length (proportional
to $\ell$); while $\xi_-$ is the length scale arising in the
hydrodynamic analysis. 
It is tempting to conclude that one has discovered the length 
scale associated with Huse and Majumdar's surface currents, but 
a more careful analysis is in order. 

\begin{figure}[tb]
\centerline{\epsfxsize=7.0cm 
\epsfbox{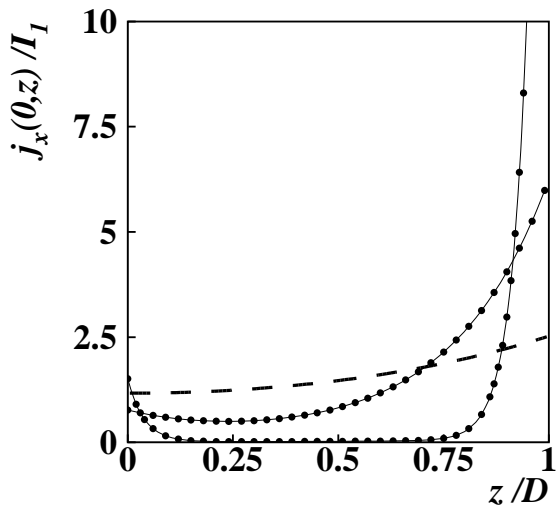}}
\vspace{1.0cm}
\caption{\label{jx}
The current density $j_x(0,z)$ for different values of $\ell$
corresponding to the Case $2$ Pad\'e conductivities,
Eqs.~(\protect{\ref{c2}}) and (\protect{\ref{small-l}}), with
parameters: $\sigma_x^0 = 5$, $\eta = 5$, $\Sigma_z = 1$, 
$D = 1$, $L = 5$ and input current $I(x)= I_1 \sin(\pi x/L)$. 
The dashed curve is the local curve ($\ell \rightarrow \infty$),
and the other curves are for $\ell=0.4$, and $\ell = 0.067$.
The areas under the curves are equal, but as $\ell$ becomes smaller
there is more current density at the surface.  }
\end{figure}

Following the steps outlined in the Appendix~\ref{pade-app} we
can calculate $V(x,z)$ for the top geometry with an input current
$I_1 \sin(\pi x/L)$.  
>From it we calculate $j_x(x,z)$, the current density in the
$x$-direction. 
Figure~\ref{jx} shows $j_x(0,z)$ as a function of $z$ for several
values of $\ell$. 
As $\ell$ decreases, the current becomes increasingly confined to 
the surface $z = D$, in other words, we obtain surface currents.
However, they differ from those found by Huse and
Majumdar\cite{huse}. 
For the same input current the hydrodynamic analysis predicts that
the ratio of current carried in the surface to that in the bulk is
\begin{equation}
\frac{\rm surface \,\,current}{ \rm bulk\,\, current} \ = \ 
\frac{\eta \pi^2}{\Sigma_z L^2};
\end{equation}
whereas Fig.~\ref{jx} seems to show that {\it all} of the current
in our solution is carried in the surface in the limit $\ell 
\rightarrow 0$. 
We confirm this result by noting that in the bulk $j_x \sim \ell$
in the small-$\ell$ limit, implying that we have no bulk and all
surface current as $ \ell \rightarrow 0$. 

We have examined the $\ell \rightarrow 0$ limit not because we
believe it to model the real conductivity but in order to make
contact with and perhaps better understand the hydrodynamic
theory. 
Nevertheless, the outcome---that much of the current is confined
to the surface---has been suggested in other contexts.
If a significant fraction of the current flows in the surface, one
might expect to find nonlinear behavior down to very low currents,
since the current density near the surface would vary quite rapidly,
leading to enormous tearing forces on the vortices. 
Nonlinear $IV$ characteristics have indeed been observed
in both the top geometry~\cite{currents} and in $\hat{c}$-axis
resistivity measurements~\cite{hellerqvist}.
However, our work is concerned only with the linear regime, and so
we do not discuss this further.

The fact that the small-$\ell$ limit of Eq.~(\ref{case2-small-l}) 
is identical to the corresponding hydrodynamic equation and
yet the two approaches predict differing amounts of surface
current suggests that it is in the treatment at the boundary (or
in the effect of the boundary layer) that the two approaches differ. 
In the Pad\'e calculation we considered the effect of the surface
only through the boundary conditions on the current; we neglected 
any effect the surface might have on the nonlocal
conductivity itself. 
One expects some surface dependence since the conductivity is
determined by the superconducting order parameter, which in turn 
depends on the boundaries.  
Within the limits of their calculation, Blum and Moore \cite{blum}
gave an explicit expression for the conductivity in the presence
of a surface.
In addition to the usual bulk conductivity their analysis yielded a
term corresponding to the image of the bulk conductivity as well
as ``cross terms."   
In order to proceed with their voltage-distribution calculation
analytically, the cross terms were dropped with an argument
suggesting their effect was small. 
Subsequent work \cite{firstyear} has shown that the effect while
small propagates farther into the bulk than was suggested in that
work, indicating again the importance of treating the surface
effects properly.  

It may turn out that the hydrodynamic approach actually incorporates
some of these surface effects. 
We have found that the small-$\ell$ limit of the Pad\'e approach
with some surface effects duplicates the hydrodynamic results. 
In the previous analysis, we used the conductivity in
Eq.~(\ref{c2}) which corresponds to
\begin{equation}
\sigma_{xx}({\bf r,r'}) = \left[\sigma_x + \frac{\eta}{\ell^2}
\right] \delta({\bf r- r'}) - \frac{\eta G(z,z')}{\ell^2}
\delta(x-x')
\label{con2}
\end{equation}
with
\begin{equation}
G(z,z') = \frac{1}{2\ell} ~{\rm e}^{-|z-z'|/\ell}.
\end{equation}
Now, we modify this conductivity so that
\begin{equation}
\label{con3}
G(z,z') = \frac{\cosh \left[ \frac{D-|z-z'|}{\ell}
\right]+\cosh \left[ \frac{D-z-z'}{\ell} \right]}
{ 2 \ell ~\sinh \left[D/\ell \right]}, 
\end{equation}
which corresponds to including a series of image terms such that
the nonlocal conductivity satisfies the boundary condition of 
its derivative vanishing on the two surfaces $z=0$ and $z=D$. 
Some motivation for using this choice might come from the boundary
condition the order parameter itself satisfies\cite{blum};
however, the justification here is that the results match those
of the hydrodynamic approach. 

\begin{figure}[tb]
\centerline{\epsfxsize=7.0cm 
\epsfbox{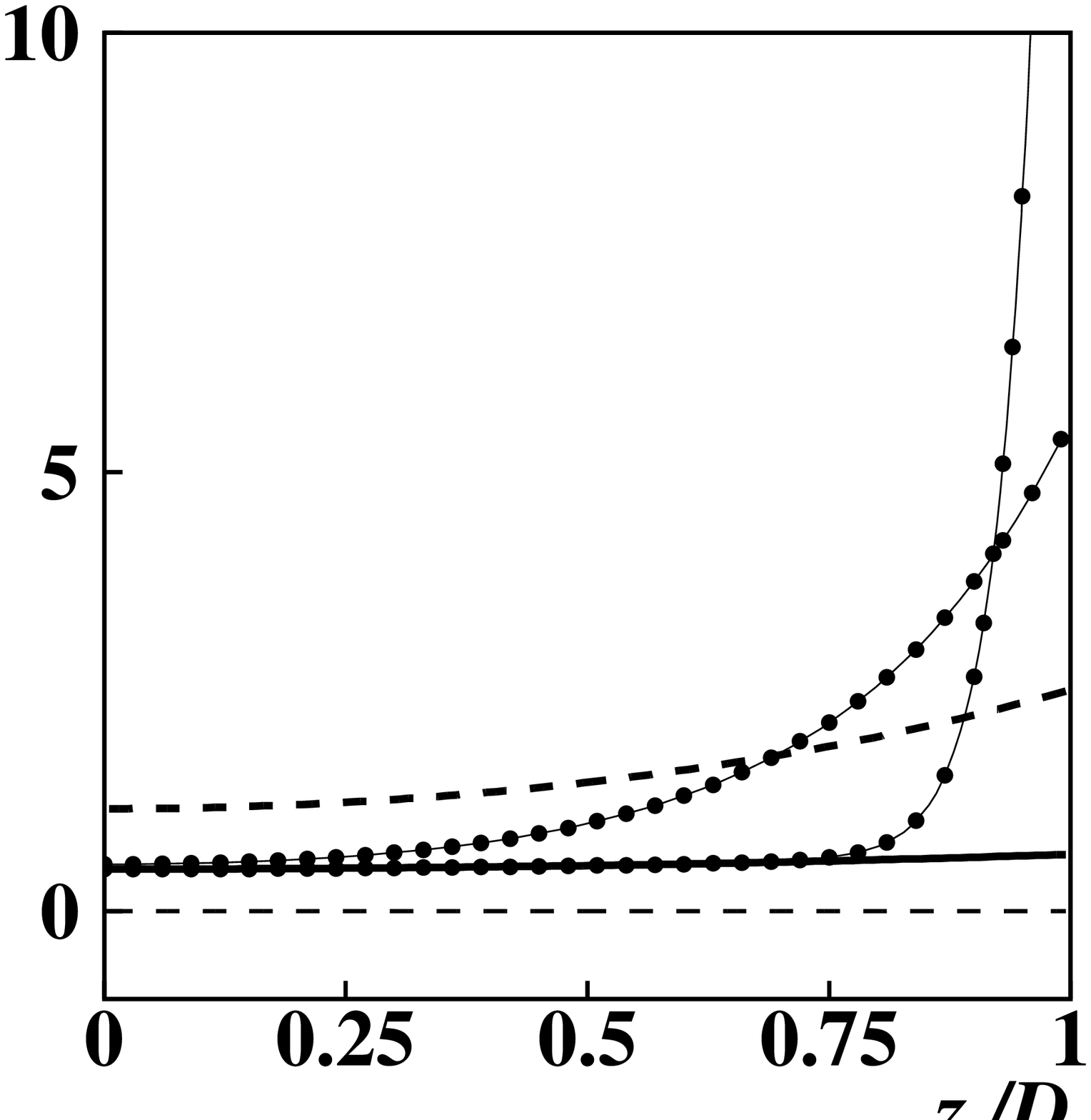}}
\vspace{1.0cm}
\caption{\label{newcond}
The current density $j_x(0,z)$ for different values of $\ell$
corresponding to nontranslationally invariant conductivities 
of the form given by
Eq.~(\protect{\ref{con3}}) with the same parameters and input
current as in Fig.~\protect\ref{jx}.
The upper dashed curve is the local curve ($\ell \rightarrow
\infty$), and the curves with $\bullet$'s are for $\ell = 0.4$
and $\ell = 0.067$. 
The lower solid curve is the Huse and Majumdar result
($j_x(0,z) \approx 0.48 \cosh(0.815 z)$ for the parameters
used here), which corresponds to the $\ell \rightarrow 0$ limit
of our results.}
\end{figure}

Using this form, one can still solve the ${\bbox \nabla} 
\cdot {\bf j} = 0$ equation in the same manner as for the
conductivity of Eqs.~(\ref{cases}).
In fact, the partial differential equation one obtains as an
intermediate step is the same as that obtained in Case $2$ with
a translationally invariant conductivity, i.e. Eq.~(\ref{fourth-2}).
But this is to be expected since the difference lies in the
boundary and not in the bulk. 
Applying the boundary conditions results in the following
voltage distribution 
\begin{eqnarray}
V(x,z)&=& \frac{I_1 \sin \left(\frac{\pi x}{L}\right)
\xi_+^2 \xi_- \left(\xi_-^2-\ell^2 \right) \cosh
\left(\frac{z}{\xi_-} \right)}
{\Sigma_z \ell^2 \left(\xi_+^2-\xi_-^2 \right)
\sinh \left(\frac{D}{\xi_-} \right)}
\nonumber \\
&&+ \left\{\xi_- \leftrightarrow \xi_+ \right\} ,
\end{eqnarray}
which agrees with the hydrodynamic result in the $\ell
\rightarrow 0$ limit. 
The corresponding current distribution $j_x(x,z)$
\begin{eqnarray}
j_x(x,z)&=& \frac{I_1 \frac{\pi}{L} \cos(\frac{\pi x}{L})
\xi_+^2 \xi_-(\eta - \sigma_x^0\xi_-^2+\sigma_x^0\ell^2)
\cosh (\frac{z}{\xi_-})}{\Sigma_z \ell^2(\xi_+^2-\xi_-^2)
\sinh(\frac{D}{\xi_-})} \nonumber \\
&&+ \{\xi_- \leftrightarrow \xi_+ \} ,
\end{eqnarray}
is plotted in Fig.~\ref{newcond}.
It can be seen that once again we have surface currents in the limit
$\ell \rightarrow 0$; however, this time only some of the current
flows in the surface, with the rest still flowing in the bulk of
the sample. 
In fact, the ratio of surface current to bulk current is identical
to that from hydrodynamics.  

This result is interesting in that it suggests that Huse and
Majumdar's boundary condition may be related to surface effects
in the conductivity. 
It is surprising that these two approaches: hydrodynamics with
its translationally invariant conductivity and its slightly
unusual boundary condition (associated with the discontinuity
in the derivative of the electric field at the surface) and 
the Pad\'e approach with a non-translationally invariant
conductivity and ordinary boundary condition, can produce
the same outcome.  

\section{Conclusions} \label{conclusions}

It is clear from the above work that the problem of nonlocal
conductivity is not completely understood; however, we have
made progress in developing a calculational scheme which
appears to be more robust than what was available previously.
Firstly, with Pad\'e approximations, we can solve for
voltage distributions having either positive or negative
viscosities; whereas, hydrodynamics is limited to positive
viscosities.
Moreover, these approximations give the correct behavior
at both high- and low-{\bf k} and so may model the surface
effects better than the hydrodynamic forms.
Because measurements in the standard flux-transformer geometry
are taken at the surface, an approach that deals inadequately
with the surface has got to be considered suspect.

Since we now have the ability to investigate both positive
and negative viscosities, we can make deductions about which
sign yields the strongly nonlocal behavior seen by Safar
{\it et al.}\,\cite{safar}
We have found that modeling these effects seems to require
positive viscosities, particularly so when the nonlocal behavior
in the $x$ direction in $\sigma_{zz}$ is considered.
As discussed in Sec.~\ref{negative?}, there is mounting
evidence that $\hat \sigma_{zz}(k_z)$ may have a negative
viscosity coefficient even at low temperatures, implying
that the nonlocal behavior seen is not due to the dependence
of $\hat \sigma_{zz}$ on $k_z$.
(There are, however, other current configurations in which
this is the only dependence of the conductivity probed.
\cite{blum})

Finally we have discussed how the conductivity itself may
be affected by surfaces and have shown that removing the
translational invariance of the conductivity does not
necessarily complicate the voltage-distribution calculation.
In particular, we have shown that the Huse and Majumdar
solution corresponds to the limit of a Pad\'e-type solution
in which the conductivity lacks translation invariance.
This surprising outcome emphasizes again the importance of
the surface in the problem of nonlocal conductivity and suggests
that even getting the correct large-$k$ behavior of the bulk
conductivity may be insufficient if one has neglected surface
effects.
Fortunately, we have at least one example in which including
the surface effects did not destroy the solubility of the
Pad\'e approach.

\acknowledgements The authors would like to thank T.\ J.\ Newman
for very helpful discussions.
S.J.P. acknowledges the support of the Engineering and Physical
Science Research Council (EPSRC).
T.B. acknowledges the support of EPSRC under grant GR/K53208 and
of the National Science Foundation under grant DMR9312476.

\appendix
\section{Analysis of the Infinite-Slab geometry}
\label{inf-app}

In this appendix we consider the length scales and surface
effects that characterize $V_T(x,z)$ and $V_S(x,z)$, the voltage
distributions in the infinite-slab geometry, by considering the
pole structure of the integrals given in Eqs.~(\ref{vtop}) and
(\ref{vside}) for a number of choices of $\hat \sigma_{zz}(k_x)$.  
The $V_S$ integral, Eq.~(\ref{vside}), is done by summing the
residues associated with the zeros of $\cosh (\kappa D/2)$, given
by $\kappa(k_x)=\pm i \pi (2m+1)/D$ with $m=0,1,\ldots$; the $V_T$ 
integral is rather similar. 
For the local problem ($\hat \sigma_{zz}(k_x)=\Sigma_z$) these poles
are evenly spaced along the imaginary axis. 
In fact, with $\delta$-function-distributed input currents
\begin{mathletters}
\begin{eqnarray}
J_{T} &=& J_0 \left[ \, \delta(x-2L_c) - \delta(x+2L_c) \right] \\
J_{S} &=& - J_0 \, \delta (x+2L_c),   
\end{eqnarray}
\end{mathletters}
the resulting expressions can be resummed to give
\begin{equation}
\label{localcase}
V_{T, S}(x,z)  = {\cal V} 
~\ln \left[ \frac{\cosh \left( \displaystyle{\frac{2L_c+x}
{\lambda_x}}\right)+ \cos\left( \displaystyle{\frac{\pi z}
{D}}\right)}{ \cosh \left(\displaystyle{\frac{2L_c \mp x}
{\lambda_x}} \right)\pm \cos\left(\displaystyle{\frac{\pi z}
{D}}\right)} \right], 
\end{equation} 
where ${\cal V}=-J_0/2 \pi \sqrt{\Sigma_x \Sigma_z}$ and 
$\lambda_x=\sqrt{\Sigma_x D^2/\pi^2 \Sigma_z}$ and where the
upper signs in the denominator correspond to $V_T(x,z)$ which
is odd about $x=0$ and the lower signs correspond to $V_S(x,z)$ 
which is odd about $z=D/2$ \cite{suggestion}. 

If we consider a nonlocal part to have the hydrodynamic form    
$\hat \sigma_{zz}(k_x) = \sigma_z^0+  \eta k_x^2 $, the poles are 
located at 
\begin{equation}
k_x(m) =  \frac{ \pm i (2m+1) }
{\left[ {\tilde \lambda}_x^2 
+ \eta (2m+1)^2 / \sigma_z^0 \right]^{1/2}},  
\end{equation}
where  ${\tilde \lambda}_x =\sqrt{\Sigma_x D^2/\pi^2 \sigma_z^0 }$. 
Note that the smallest pole ($m=0$) is shifted to smaller $k_x$ 
compared to the local situation ($\eta=0$), implying a longer
length scale. 
 There is only a significant shift if the viscous length scale 
$\sqrt{\eta/\sigma_z^0}$ becomes comparable to the ``local" length 
scale $\tilde \lambda_x$ which depends on the sample thickness $D$.  
 As $\eta$ increases further the viscous length dominates; 
we were able to resum the series in this limit, finding 
\begin{equation}
V_S(x,z) \approx \frac{-J_0(2z-D)}{4 \sqrt{\sigma_z^0 \eta}} 
\exp \left\{  -\sqrt{\frac{\sigma_z^0}{\eta}} ~|2L_c+x| \right\}. 
\end{equation}
It would be interesting to probe the spatial dependence of $V_{S}$ 
experimentally and extract its length scale. 
However, such a measurement would be difficult as it would require a 
sample long enough to accommodate several leads, and the voltages far 
from the current may become too small to be meaningful. 
Returning to the pole structure in the hydrodynamic case, another 
point to notice is that they accumulate at a finite value 
$\pm i \sqrt{\sigma_z^0/\eta}$. 
As a result $V_S(x,z)$ no longer diverges logarithmically at the 
contact points $(-2L_c,D)$ and $(-2L_c,0)$. 
Given the $\delta$-function input currents, this divergence is 
physical, and the failure of the hydrodynamic form to reproduce it 
is an example of how the incorrect large-${\bf k}$ can affect the 
potential especially at the surface.  

Next, let us consider $\hat \sigma_{zz}(k_x)$ to have a  Pad\'e form
\begin{equation}
\label{pade-1}
\hat \sigma_{zz}(k_x) = \Sigma_z  +
\frac{\Delta_z^x}{1+ k_x^2 \ell^2}. 
\end{equation} 
The small-$k_x$ behavior of this expression is similar to the 
hydrodynamic example if $\Delta_z^x < 0 $.  
However, it has twice as many poles since $\kappa(k_x)=\pm i \pi 
(2m+1)/D$ has twice as many solutions as in the hydrodynamic
case. 
These poles break into two sets. 
For large $m$ one set is evenly spaced and mimics the behavior in 
the local problem including the logarithmic divergence at the leads,
while the other set accumulates at the value $\pm i
\sqrt{(1+\gamma_z^x)/\ell^2}$ (where $\gamma_z^x=\Delta_z^x/\Sigma_z$)
and mimics the hydrodynamic behavior in the bulk.  
The Pad\'e form has clear advantages over the hydrodynamic form, but 
even it is not quite right since the large-$k_x$ limit of $\hat 
\sigma_{zz}(k_x)$ should be less than $\hat \sigma_{zz}(0)$. 
To achieve that one needs something like
\begin{equation}
\hat \sigma_{zz}(k_x) = \Sigma_z \left[1+\frac{\gamma_1}{1+k_x^2
\ell^2} -\frac{\gamma_2} {(1+k_x^2 \ell^2)^2} \right]
\end{equation}
with $\gamma_1/2 \leq \gamma_2 \leq \gamma_1$, which leads to three 
sets of poles---one like the local case and two like the 
hydrodynamic case. 
With the Pad\'e form, Eq.~(\ref{pade-1}), one can also investigate 
the consequences of ``negative" viscosities when $\Delta_z^x > 0$.  
As $\Delta_z^x$ increases, the $x$-axis length scale decreases and 
eventually a point is reached at which the poles move off the 
purely imaginary axis and some oscillatory behavior is superimposed 
on the exponential decay.    


\section{Solution of the Integro-differential equation 
associated with the Pad\'e form}
\label{pade-app}

In this appendix we outline the solution of Case $2$. 
For conductivities of the form given by Eq.~(\ref{c2}),
the steady-state continuity equation ${\bbox \nabla} \cdot
{\bf j} = 0$ takes the form 
\begin{eqnarray}
\label{divj}
&& \Sigma_x \partial_x^2 V(x,z) 
+ \Sigma_z \partial_z^2 V(x,z) \nonumber \\ 
&& + \frac{\Delta_x^z}{2 \ell}  
\left[\int_0^D {\rm e}^{-|z -z'|/\ell} \, \partial_x^2 V(x,z')
\, dz' \right] =0, 
\end{eqnarray}
where ${\rm exp}\{-|z-z'|/\ell\}/2\ell$ is the Fourier transform
of $(1+k_z^2 \ell^2)^{-1}$.

Let us now apply the following trick. 
 Differentiate Eq.~(\ref{divj}) twice with respect to $z$, 
which leads to 
\begin{eqnarray}
\label{trick}
\Sigma_x \partial_z^2 \partial_x^2 V(x,z)
+\Sigma_z \partial_z^4 V(x,z) 
- \frac{\Delta_x^z}{\ell^2} \partial_x^2 V(x,z) \nonumber \\ 
+ \frac{\Delta_x^z}{2 \ell^3} 
\left\{\int_0^D e^{-|z -z'|/\ell} \partial_x^2 V(x,z') dz' 
\right\} =0, 
\end{eqnarray}
where we have exploited the relation 
\begin{equation}
\frac{d^2}{dz^2}\left({\rm e}^{- |z-z'|/\ell }\right) =
\left[ \frac{1}{\ell^2} - \frac{2}{\ell} \delta (z-z') \right]  
{\rm e}^{- |z-z'| / \ell },
\end{equation}
familiar from the solution of Schr\"odinger's equation with a 
$\delta$-function potential. 
We can combine Eqs.~(\ref{divj}) and (\ref{trick}) to 
eliminate the integral term, giving 
\begin{equation}
\label{fourth}
\left[\Sigma_z\partial_z^4 + \Sigma_x
\partial_z^2 \partial_x^2 
-\frac{\left(\Sigma_x+\Delta_x^z \right)}{\ell^2} \partial_x^2   
-\frac{\Sigma_z}{\ell^2} \partial_z^2  \right] V=0,
\end{equation}
which was given in the main body of the paper as
Eq.~(\ref{fourth-2}). 

Separating variables, $V(x,z) = X(x) \, Z(z)$, and applying
the boundary condition $j_x(\pm L/2,z)=0$   yields 
$X(x)=A\cos(2 n \pi x/L)$ or $X(x)=B\sin[(2n+1) \pi x/L]$
for $n=0,1,\ldots$. 
When $n > 0$, the corresponding $Z(z)$ is given by  
\begin{eqnarray}
\label{formofsoln}
Z_n(z) =&& 
P_n \,\cosh \left(\frac{z-\frac{D}{2}}{\xi_+}\right) + 
Q_n \,\sinh \left(\frac{z-\frac{D}{2}}{\xi_+}\right) 
\nonumber \\+&& 
R_n \,\cosh \left(\frac{z-\frac{D}{2}}{\xi_-}\right) + 
S_n \,\sinh \left(\frac{z-\frac{D}{2}}{\xi_-}\right),
\end{eqnarray}
where $\xi_{\pm} \equiv \xi_{\pm}(n)$ is given by Eq.~(\ref{omega2}).
For some of the algebra that follows it is convenient to choose
modes that are symmetric about $z=D/2$. 

It might appear that we need to apply four boundary conditions to 
determine the constants in $Z_n(z)$, but actually two of the 
constants are found by inserting the solution into the original 
integro-differential equation, Eq.~(\ref{divj}). 
This step yields the following two conditions 
\begin{mathletters}
\begin{eqnarray}
\frac{P_n}{R_n}&=& \frac{-\xi_-(\ell^2-\xi_+^2)
\left[\xi_- \cosh(\frac{D}{2 \xi_-})+
\ell \sinh(\frac{D}{2\xi_-}) \right]}
{\xi_+(\ell^2-\xi_-^2)
\left[\xi_+ \cosh(\frac{D}{2 \xi_+})+
\ell \sinh(\frac{D}{2\xi_+}) \right]},
\\
\frac{Q_n}{S_n}&=& \frac{-\xi_-(\ell^2-\xi_+^2)
\left[\ell \cosh(\frac{D}{2 \xi_-})+
\xi_- \sinh(\frac{D}{2\xi_-}) \right]}
{\xi_+(\ell^2-\xi_-^2)
\left[\ell \cosh(\frac{D}{2 \xi_+})+
\xi_+ \sinh(\frac{D}{2\xi_+}) \right]}.
\end{eqnarray}
\end{mathletters}
The two remaining constants are fixed by the boundary conditions
on $j_z$, namely  
\begin{eqnarray}
\label{boundary2}
-\Sigma_z \partial_z V(x,D)  &=& I_{top}(x),
\nonumber \\ 
-\Sigma_z \partial_z V(x, 0) &=& I_{bot}(x).
\end{eqnarray}
They turn out to be fairly complicated functions of the parameters 
$\ell$, $\Sigma_x$, $\Sigma_z$, $\gamma_z^z$, $n$ and $L$, in 
addition to the Fourier components of $I_{top}(x)$ and $I_{bot}(x)$. 
 Note that each $Z_n(z)$ is only a function of the 
corresponding Fourier component of the current. 
 The $n=0$ part of the solution requires separate consideration;
however, it is straightforward and no details are provided
here.

\section{Bose glass scaling forms for conductivities}
\label{bose}

In this section we consider scaling forms for the conductivities 
which might be expected to hold in the presence of correlated
disorder ({\it e.g.}\ columnar defects or twin boundaries). 
With such disorder, there is thought to be a second-order phase
transition between the low-temperature Bose-glass phase and the
high-temperature phase consisting of an entangled liquid of
delocalized flux lines \cite{nelson1,wallin1}. 
Near the transition the characteristic length scales $\ell_{\perp}$
(within the $ab$ planes) and $\ell_{\parallel}$ (along the
$\hat{c}$-axis) and the characteristic time scale $\tau \sim
\ell^{z'}_{\perp}$ diverge.
Nelson and Radzihovsky\cite{nelson3} used the scaling of the
free-energy density, $f \sim 1/\ell_{\parallel} \ell_{\perp}^2$,
and the vector potential, $A_{\parallel} \sim 1/\ell_{\parallel}$
and $A_{\perp} \sim 1/\ell_{\perp}$ (from gauge invariance), and
the relations ${\bf J}=\partial f/\partial {\bf A}$ and ${\bf E}=
-\partial {\bf A}/ \partial t$ to suggest that the conductivities
scale as follow
\begin{eqnarray}
\sigma_\perp &\sim& \ell_\parallel^{-1} ~\ell^{z'}_\perp,
\nonumber \\
\sigma_\parallel &\sim& \ell_\parallel ~\ell^{z'-2}_\perp .
\end{eqnarray}
Studies of this transition\cite{wallin1,nelson3} have suggested
that  $\ell_\parallel \sim\ell_\perp^2 $ and $z' = 6.0 \pm 0.5$;
we are going to use $z' = 6$. 

In the Pad\'e-form conductivities (Eqs.~\ref{cases}), we have
a length scale $\ell$, which is a $\hat{c}$-axis length scale
$\ell_\parallel$ in Cases $1$ and $2$ and is an $ab$-plane length
scale $\ell_\perp$ in Cases $3$ and $4$. 
Recall the conductivities in Case $1$ are $\hat{\sigma}_{xx}({\bf k})
 = \Sigma_x $ and $\hat{\sigma}_{zz}({\bf k}) = \Sigma_z +
\Delta_z^z/(1 + k_z^2 \ell^2)$.
What we want to do here is determine the dependence of the constants 
$\Sigma$ and $\Delta$ upon $\ell$. 
Since in this case the length scale is $\ell_{\parallel}$, we use
$\ell_\parallel \sim \ell_\perp^2$ to eliminate the dependence
upon $\ell_\perp$ and arrive at 
\begin{eqnarray}
\sigma_\perp^{(s)} &\sim& \ell^2_\parallel ,
\nonumber \\
\sigma_\parallel^{(s)} &\sim& \ell^3_\parallel
~f(k_z \ell_{\parallel}),
\end{eqnarray}
where the superscript $(s)$ refers to the superconducting
contribution, we also include in $\Sigma_x$ and $\Sigma_z$ normal
contributions $\sigma_x^{(n)}$ and $\sigma_z^{(n)}$ that are not
affected by the scaling. 

We thus obtain the following Bose-glass scaling forms:
\begin{mathletters}
\begin{eqnarray}
\label{case1sca}
1. \ \ 
\hat{\sigma}_{xx}({\bf k})&=& \sigma_x^{(n)} +
 C_2 \ell_{\parallel}^2, \nonumber \\
\hat{\sigma}_{zz}({\bf k})&=&
\sigma_z^{(n)} + \frac{ C_1 \ell_{\parallel}^3}
{ 1+ k_z^2 \ell_{\parallel}^2}; \\
&& \nonumber \\
\label{case2sca}
2. \ \ 
\hat{\sigma}_{xx}({\bf k}) &=& \sigma_x^{(n)} +
\frac{C_2 \ell_{\parallel}^2}{1 + k_z^2
\ell_{\parallel}^2}, \nonumber \\
\hat{\sigma}_{zz}({\bf k}) &=& \sigma_z^{(n)} +
C_1 \ell_{\parallel}^3; \\
&& \nonumber \\
\label{case3sca}
3. \ \ 
\hat{\sigma}_{xx}({\bf k}) &=& \sigma_x^{(n)} +
\frac{C_2 \ell_{\perp}^4}{1 + k_x^2 \ell_{\perp}^2},
\nonumber \\
\hat{\sigma}_{zz}({\bf k}) &=& \sigma_z^{(n)}
+ C_1 \ell_{\perp}^6; \\
&& \nonumber \\
\label{case4sca}
4. \ \  
\hat{\sigma}_{xx}({\bf k}) &=&  \sigma_x^{(n)} +
C_2 \ell_{\perp}^4, \nonumber \\
\hat{\sigma}_{zz}({\bf k}) &=& \sigma_z^{(n)} +
\frac{C_1 \ell_{\perp}^6}{1 + k_x^2 \ell_{\perp}^2}.
\end{eqnarray}
\end{mathletters}
It should be noted that although the constants $C_1$ and $C_2$
have no explicit dependence upon the length scales, they will be
temperature-dependent.
However, compared to the temperature dependence of the length
scales $\ell_\perp$ and $\ell_\parallel$ near the transition,
which go as powers of $\vert T - T_{BG} \vert$ (where $T_{BG}$ is
the transition temperature), it is a weak dependence.
In the forms above the coefficient of the Pad\'e term is
assumed to be positive, and therefore the viscosity coefficient
is negative.
We can also write forms that have positive viscosity coefficients; 
for instance, Case $4$ would be 
\begin{eqnarray}
\label{case4sca-pos} 
\hat{\sigma}_{xx}({\bf k}) &=& \sigma_x^{(n)} + C_2 \ell_{\perp}^4,
\nonumber \\ 
\hat{\sigma}_{zz}({\bf k}) &=& \sigma_z^{(n)} + 2 C_1 \ell_{\perp}^6 
-\frac{C_1 \ell_{\perp}^6}{1 + k_x^2 \ell_{\perp}^2}.
\end{eqnarray}
which is the form used to generate Fig.~\ref{scaled}. 


\widetext

\begin{thebibliography}{999}
\bibitem{bc} We also find the Pad\'e boundary conditions at the 
sides $\pm L/2$ reduce those of Huse and Majumdar, $\left(\sigma_x^0 
\partial_x - \eta \partial_x \partial_z^2 \right) V(\pm L/2,z)$ in 
the $\ell \rightarrow 0$ limit provided we are away from $z=0$ 
and $z=D$.
\end{thebibliography}
\end{document}